\documentstyle[12pt]{article}
\newcommand{\delslash}{\not\!\partial}

\begin{document}
\begin{flushright}
NIIG-DP-96-2\\
October, 1996
\end{flushright}
\begin{center}
{\Large Euclidean Solutions in Broken Phase\\
 and Electro-Weak Dynamics}\\
\vskip .75in

{\large Hiroyuki Kanada, Hiroto So and Shin Takeda}\\
\vskip .2in
{\it Department of Physics},\\
{\it Niigata University},\\
{\it Ikarashi 2-8050,
 Niigata 950-21, Japan.}\\

\vskip .5in
\end{center}
\begin{abstract}
\baselineskip 16pt
A Higgs-Yukawa system in a broken phase and Euclidean solutions are
  investigated.
Although it has been believed that there are no Euclidean solutions
in the broken phase in 4-dimension,  
we find numerically ones in the phase due to the effect of 
 a strong Yukawa coupling. 
 The complex Yukawa coupling is necessary for 
 the stability of the solution.
 The extension to a complex Higgs-Yukawa system is also investigated.
\end{abstract}
\setcounter{footnote}{0}

\newpage
\baselineskip 18pt

{\it 1.\ Introduction\ \ }
In the electro-weak (EW) theory with rich structure of the dynamics, 
it is an essential point that the Higgs field has a non-zero 
vacuum expectation value (VEV).
  Its typical example is the existence of static solutions as sphaleron
which is unstable and combine distinct vacua from each other in the  EW
theory. 
  It is important that there is an instanton solution in the system to
estimate the tunneling effect between distinct vacua semi-classically. 
  On the other hand, by the famous scaling argument \cite{thft,aflk}, it
is pointed out that the EW theory can not have instanton solutions
because of collapsing phenomena.

  To solve  the above problem we know a constrained
instanton, a streamline method and new valley one [2-4].
  They are not exact solutions in the theory but an effective
collection of configurations in the path integral.  
  Another approach is to introduce some external forces against the
collapse.
  One possibility is given by taking account of the effect of heavy
fermions.  
  One of the evidences is the fermion effect on sphaleron
[5-9].
 The sphaleron barrier between two vacua is more lowered by  heavier
fermion effects. 
 This means the transition between distinct vacua becomes to be easier.

  Now, we can see the naive scaling argument as the following:
  For the Higgs field, we can give a scale transformation with fixing
its VEV as 

$$
\phi(x,a) \rightarrow \phi(ax).
$$

 In addition to spatial coordinates, other
fields are transformed like as
canonical scaling. Then the scaling of the present action  is  given as
follows; 
$$
S(a) = S_0 + S_{-4}a^{-4} + S_{-2}a^{-2} + S_{-1}a^{-1},
$$
where $S_0$ is a scale invariant term such as gauge and matter kinetic
ones without VEV.  
  The coefficient $S_{-4}$ is a positive number due to giving a
stability of the vacuum and the coefficient $S_{-2}$ is a positive
number which 
comes from the  kinetic  term of the scalar field with VEV in the
Lagrangian. On the other hand, the sign of the coefficient $S_{-1}$ can
not be definitely determined because of the Yukawa coupling effect. 
 
 From the above scaling argument, one can understand that 
the Yukawa term ($S_{-1}$) plays an important role in the short distance
phenomena. 
  If the term is negative and has enough magnitude, the collapsing of
the solution becomes to be impossible by the repulsing effect. 
 
  In this article, we investigate a Higgs-Yukawa system in a broken
phase and construct Euclidean solutions (instantons) numerically for a
strong Yukawa coupling.

{\it 2. \ simple Higgs-Yukawa System\ \ }
  We shall derive equations of motion for a simple Higgs-Yukawa system
 (the Higgs filed is real) in a Euclidean space.
  Since fermi fields $\Psi$ and $\bar{\Psi}$ are independent of each
other in the Euclidean space, the real positiveness of $\bar{\Psi} \Psi$ is
not guaranteed. 
  So, we can introduce a complex Yukawa coupling $g(=|g| \mbox{e}^{\mbox{i}
 \theta})$, keeping a Lagrangian hermite. Here, the Lagrangian is given as

\begin{equation}
\label{eq:lagrangian}
 {\cal L} = \frac{1}{2}(\partial_\mu \Phi)^2 
  + 2 \lambda^2( \Phi^2 -  v^2)^2
  + \bar{\Psi} \gamma \cdot \partial \Psi
  + \Phi \mbox{Re}(g \bar{\Psi} \Psi ),
\end{equation}
where $v$ is the VEV of the Higgs field $\Phi$ and is taken as
$v=246$ GeV which correspond to a standard Higgs in the EW theory.
  By introducing the spherical ansatz for the scalar field, one can be
applicable for simplicity as
\[
\Phi (x) = \Phi (r),
\]
where $r=\sqrt{x^2}$.

The ansatz for the fermi fields is non trivial because of taking account
of  freedom of spinor. Here, we adopt the following ones;

\begin{eqnarray}
\nonumber
\Psi(x) &= & (f_1 (r) + \mbox{e}^{\mbox{i} \theta} f_2 (r) \gamma_{\mu}
\frac{x_{\mu}}{r})\xi,\\ 
\nonumber
\bar{\Psi}(x) &=& \bar{\xi}(f_1 (r) - \mbox{e}^{\mbox{i} \theta} f_2 (r)
\gamma_{\mu} \frac{x_{\mu}}{r}),
\end{eqnarray}
where $\xi$ and $\bar{\xi}$ denote constant spinors. The
normalization is given as
\begin{eqnarray}
\nonumber
\bar{\xi} \xi &=& \pm \mbox{e}^{- \mbox{i} \theta},\\
\nonumber
\bar{\xi} \gamma_{\mu} \xi &=& 0.
\end{eqnarray}

Unfortunately, there are no explicit forms of the equations of motion for
$\Psi$ and $\bar{\Psi}$ except for the $\theta = 0$ or $\frac{\pi}{2}$
case. 
Thus, by substituting the 
above ansatz into the Lagrangian (\ref{eq:lagrangian}), one can derive  the
fermion equations of motion in term of $f_1$ and $f_2$.
So, we can write  down equations of motion by $\Phi$, $f_1$ and $f_2$ as
\begin{eqnarray}
\label{eom:phi}
\Phi'' + \frac{3}{r} \Phi' &=& 8\lambda^2\Phi(\Phi^2-v^2) \pm |g|
 (f^2_1-  f^2_2\cos 2\theta),\\
\label{eom:f1}
f'_1 &=& |g| \Phi f_2\cos 2\theta,\\
\label{eom:f2}
f'_2 + \frac{3}{r}f_2 &=& |g| \Phi f_1.
\end{eqnarray}
  Here, one can see that in the $\theta = 0$ or $\frac{\pi}{2}$ case,
above equations are reduced ones which are derived from equations of  
motion of $\Psi$ and $\bar{\Psi}$. 
  Particularly, the Yukawa coupling with $\theta=0$ becomes to be real.
Thus both $f_1$ and $f_2$ have real masses and then have normal and
exponential dumping behaviors.
  For the exponential behavior of $\Phi$, the Eq.(\ref{eom:phi})
indicates that the fermion mass must be heavier than a half of the Higgs 
mass, thus the Yukawa coupling must be sufficiently strong.
  Among  present known fermions, only top quark may be satisfied with
this condition, so we expect that top quark plays a main role for an
instanton of the Higgs-Yukawa system. 

We examine Eq.(\ref{eom:phi})\ -\ Eq.(\ref{eom:f2}) numerically to 
find a solution that obeys the following boundary conditions;

\[
\Phi'(0) = f_2(0) = 0,
\]
\[
\Phi(\infty) = v,
\]
\[
f_1(\infty) = f_2(\infty) = 0.
\]
  The first condition is given for the smoothness at the origin, the
second one means the vacuum condition at $r=\infty$. The third one gives 
the  normalizability condition of fermionic wave function.  
  In the calculation, both values of $f_1(0)$ and $\Phi(0)$ are free
parameters. So, we treat the $f_1(0)$ as a variable parameter with a
fixed $\Phi(0)$ to find a solution that satisfies the above boundary
conditions. 

  Before numerical investigation of the Higgs-Yukawa system, we examine
a pure Higgs system analytically.
  In the pure Higgs system, there are two kinds of  ``solutions'' that
do not obey boundary conditions. Eq.(\ref{eom:phi}) without Yukawa term
is interpreted as the point particle's equation of motions with a
friction 
in a potential of $-(\Phi^2-v^2)^2$ \cite{coleman}.  In a first type
solution with $|\Phi(0)|<v$, the particle falls into the origin and
oscillates  dumpingly around the origin.  
  In  a second type one with $|\Phi(0)|>v$, the particle departs from
the 
origin and the ``solution'' behaves asymptotically  as $\Phi(r) \propto
\frac{1}{a+r}$ where $a$ is a constant.
  Then, we can determine a sign of Yukawa coupling to improve the above
behavior.  

Now, we try to search the numerical solution in the special three cases
as  
$\theta=0$ (a real Yukawa coupling), 
$\theta=\frac{\pi}{2}$ (a imaginary Yukawa coupling) and 
$\theta=\frac{\pi}{4}$.
 
 In the case of $\theta=0$, we can not find any solutions which obey the
boundary conditions for any $\Phi(0)$.  
  However, we can find ``solutions'' very similar to one in the pure
Higgs system.
  It is pointed out that when $\Phi(r),f_1(r),f_2(r) >> 0 $,
Eq.(\ref{eom:phi})\ -\ Eq.(\ref{eom:f2}) have simple pole solutions as
\[
 \Phi(r) \propto  f_1(r) \propto f_2(r)
 \sim \frac{1}{a+r}.
\]
The result for $\Phi(r)$ obtained  for many values of $f_1(0)$
 are given in Fig.1. One can see in this figure
that all calculated  $\Phi(r)$ show a simple pole behavior in
 all values of $f_1(0)$.

Next, we try to search for a solution in the case of 
$\theta=\frac{\pi}{2}$ .
We can get many numerical 
solutions which obey the boundary conditions.  
  The typical results for $ \Phi(r) , f_1(r)$ and $ f_2(r)$ are shown in
Fig.2\ -\ Fig.4, respectively.  
  These figures show that these solutions localizes around the origin with
nonzero finite size. It means that the contribution from fermion 
supports for the Higgs field not to collapse to the trivial 
solution. 
  It has already be pointed out that there is a analytic solution in
this theory with $v=0$ by Inagaki \cite{inagaki}.
  The solution is shown with a dashed line in Fig.2\ -\ Fig.4, too.  

In the case of $\theta=\frac{\pi}{4}$, only $f_1$ contributes to Higgs
field, 
and $f_1$ is linear in $r$. 
  Thus, one can easily know that there are no solutions. 
  This fact means that there exists a gap between two regions connected
to $\theta=0$ and to $\theta=\frac{\pi}{2}$.

  We investigate numerically whether there are solutions in the present
system or not, in the two cases of  $0 \le \theta < \frac{\pi}{4}$ and
$\frac{\pi}{4} < \theta \le \frac{\pi}{2}$.  
  We find that there is not any solution in the case of $0 \le \theta <
\frac{\pi}{4}$.  
  On the other hand, there are solutions for $\frac{\pi}{4} < \theta \le
\frac{\pi}{2}$.
  To show these facts, the actions calculated with the obtained
solution is given as a function of $\theta$ in Fig.5.

  We can show also the relation of imaginary fermion masses to the
calculated action in Fig.6. 
  One can see that the mass dependency of the action is consistent with
the result given by Nolte and Kunz \cite{kunz2}.

Finally, we can point out that important one of present findings is that
the solutions localize around the origin with a nonzero finite size, and the
contribution from fermions plays an important role for the Higgs field
not to 
collapse to the trivial solution.

{\it 3.\ Complex Higgs-Yukawa System\ \ }
  Next, we try the extension to  complex Higgs-Yukawa system.  We
introduce two same fermions except for constant spinors normalization to
avoid a linear term of $x$ in Lagrangian with a suitable ansatz for
fields.
  Thus, our Lagrangian is,
\[
{\cal L} = |\partial \Phi|^2  
               + 2 \lambda ( |\Phi|^2 - v^2)^2
               + \bar{\Psi} \delslash \Psi 
	       + \bar{\Psi}_{L} \Phi \Psi_{\mbox{\tiny R}}
	       + \bar{\Psi}_{\mbox{\tiny R}} \Phi^{\dagger} \Psi_L,
\]
where $\Psi = {}^t\left[ \psi_1,\psi_2 \right]$ and $\Phi =
\Phi_{\mbox{\tiny R}} + i \Phi_{\mbox{\tiny I}}$.  

In the present investigation, the ansatz for each field is given the
following,
\begin{eqnarray}
\nonumber
\Phi &=& \Phi(r),\\
\nonumber
\psi_i &=& (f_1 + f_2 (\gamma \cdot \hat{x}) 
          + \mbox{i} f_3 \gamma_5 
          + \mbox{i} f_4 \gamma_5 (\gamma \cdot \hat{x})) \omega_i, \\
\nonumber
\bar{\psi}_i & =& \bar{\omega}_i(f_1 - f_2 (\gamma \cdot \hat{x}) 
          + \mbox{i} f_3 \gamma_5 
          - \mbox{i} f_4 (\gamma \cdot \hat{x}) \gamma_5),\\
\nonumber
\bar{\omega}_i \omega_i                     &=& A \ \ \mbox{(real number)},\\
\nonumber
\bar{\omega}_i \gamma_5 \omega_i            &=& B \ \ \mbox{(imaginary
number)},\\ 
\nonumber
\bar{\omega}_i \gamma_5 \gamma_\mu \omega_i 
     &=& \left\{ 
         \begin{array}{cll}
            C  & \mbox{(imaginary number)} \mbox{\ \ for\ } i = 1, \\
            -C & \mbox{(imaginary number)} \mbox{\ \ for\ } i = 2.
         \end{array}
         \right.
\end{eqnarray}

Finally, we can obtain following equations of motion, using the above ansatz.
\begin{eqnarray}
\label{eq:cHY-PHIR}
\Phi_R''(r) + \frac{3}{r} \Phi_R'(r)
&=&
4 \lambda \Phi_{\mbox{\tiny R}}( |\Phi|^2 - v^2) \\
\nonumber
&& + g A (f_1^2 - f_2^2 - f_3^2 + f_4^2)
+ 2 \mbox{i} g B (f_1 f_3 + f_2 f_4 ),\\
 \Phi_I''(r) + \frac{3}{r} \Phi_I'(r)
&=&
 4 \lambda \Phi_{\mbox{\tiny I}}( |\Phi|^2 - v^2) \\
&&
\nonumber
+ \mbox{i} g B (f_1^2 + f_2^2 - f_3^2 - f_4^2 )
 - 2 g A (f_1 f_3 - f_2 f_4),\\
f_1' &=& g ( \Phi_{\mbox{\tiny R}} f_2 - \Phi_{\mbox{\tiny I}} f_4),\\
f_2' + \frac{3}{r} f_2 &=& g ( \Phi_{\mbox{\tiny R}} f_1 
- \Phi_{\mbox{\tiny I}}
 f_3),\\
f_3' &=& - g ( \Phi_{\mbox{\tiny R}} f_4 + \Phi_{\mbox{\tiny I}} f_2),\\
\label{eq:cHY-F4}
f_4' + \frac{3}{r} f_4 &=& 
- g ( \Phi_{\mbox{\tiny R}} f_3 + \Phi_{\mbox{\tiny I
}} f_1).
\end{eqnarray}
Here, boundary conditions are as follows,
\[
\Phi_{\mbox{\tiny R}}'(0) = \Phi_{\mbox{\tiny I}}'(0) = f_2(0) = f_4(0) =0,
\]
\[
\Phi_{\mbox{\tiny R}}(\infty)^2 + \Phi_{\mbox{\tiny I}}(\infty)^2 = v^2,
\]
\[
 f_1(\infty) = f_2(\infty) =  f_3(\infty) = f_4(\infty) = 0.
\]
  
The first condition is given by  the smoothness at the origin as same as
the one in simple Higgs case. 
the second one means vacuum at infinity and the third one denotes
fermion normalizability.  

We shall examine equations of motion in this system in comparison 
with those in the previous simple Higgs-Yukawa system.
  Putting $\Phi=\Phi_{R}$ , $f_3=f_4=0$ and $B=0$ in
Eq.(\ref{eq:cHY-PHIR})\ -\ Eq.(\ref{eq:cHY-F4}), we can obtain the same
equations in the simple Higgs-Yukawa system with $\theta = 0$.
  Then one can know that there are simple pole solutions in the present
system.
  If we take as $f_2\rightarrow\mbox{i} f_2$ , $g\rightarrow \mbox{i} g$
and 
$f_3=f_4=0=B=0$  in Eq.(\ref{eq:cHY-PHIR})\ -\ Eq.(\ref{eq:cHY-F4}), we
can get same equations
in the previous system with $\theta=\frac{\pi}{2}$.
 In this case, we can say that there is a solution with a nonzero finite
size. 

Now, we are going to search for a solution with a nonzero finite size in
the present system.
  It is important for constructing instanton solution in the EW theory
to get a solution in this system.

{\it 4.\ Summary\ \ }
  We shall summarize this article.
Due to the strong Yukawa coupling, the Euclidean solution in a broken
phase is stabilized. 
Explicit solutions are found numerically.
 Fermi fields can give  the effect as external sources against
the collapsing instanton. 
  Generally, external sources enough to support instantons must have
suitable values or sign in the following three points; (1)magnitude,
(2)spatial size and (3)sign.

It is so difficult to change the short distance behavior in the standard
dynamics  except Yukawa couplings, and then we can not find the other
possibility for external sources against collapse.

Changing of the Yukawa coupling from real to imaginary corresponds to
the above third point and prefer repulsive effect a to attractive ones
owing to giving against collapsing forces.

  Our final aim is to construct instanton solutions in the EW theory.
  In the theory, there are other problems gauge fields, index theorem
and fermion zero modes. 
  They shall be studied in the succeeding paper.

\newpage
\pagestyle{empty}
{\large Figure Captions}

\begin{itemize}
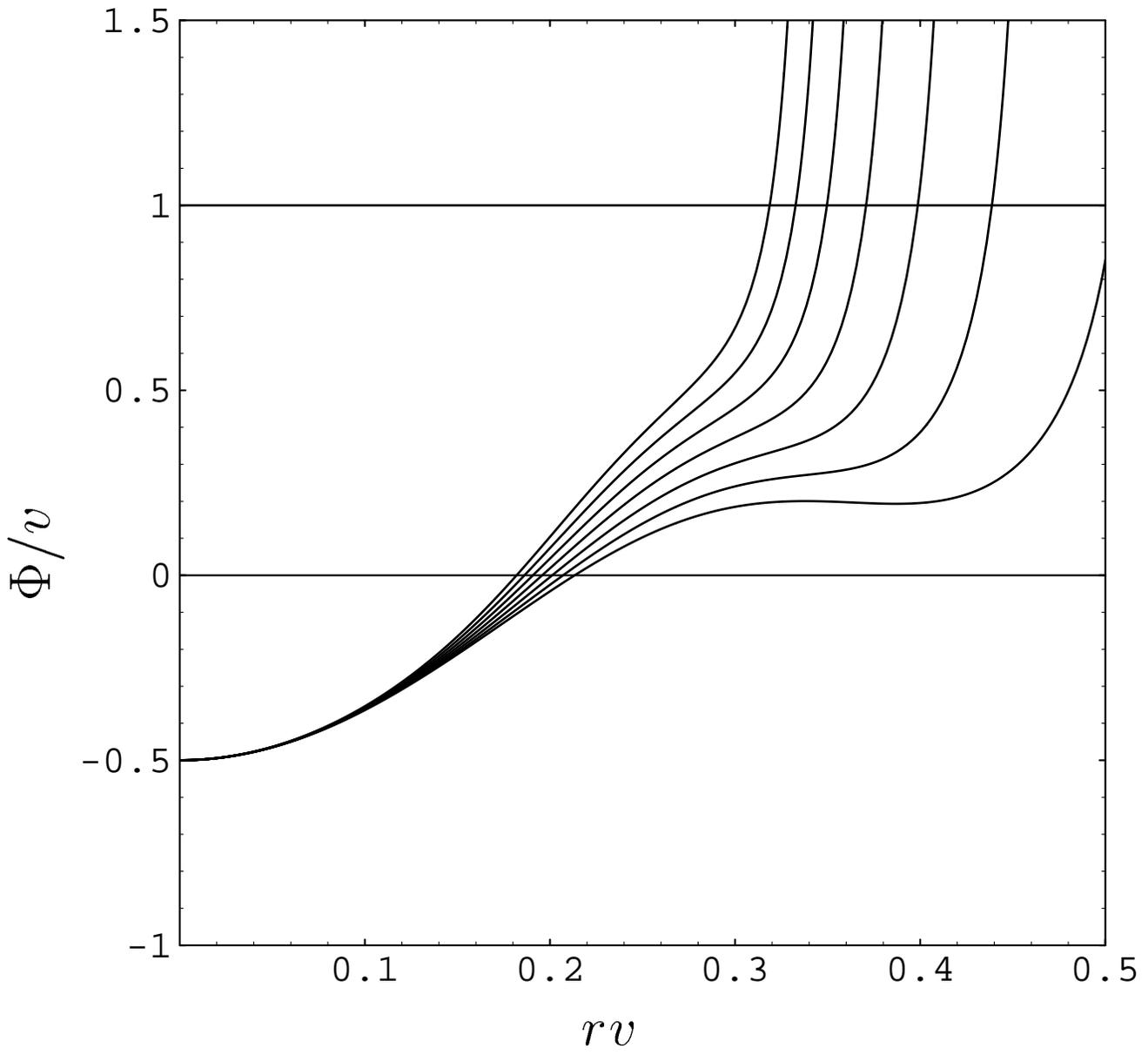
 \item{Fig.1}
      ~~ Numerical ``Solutions'' for the Higgs-field in the real Yukawa
      system. Each line denotes one for different values of $f_1(0)$.
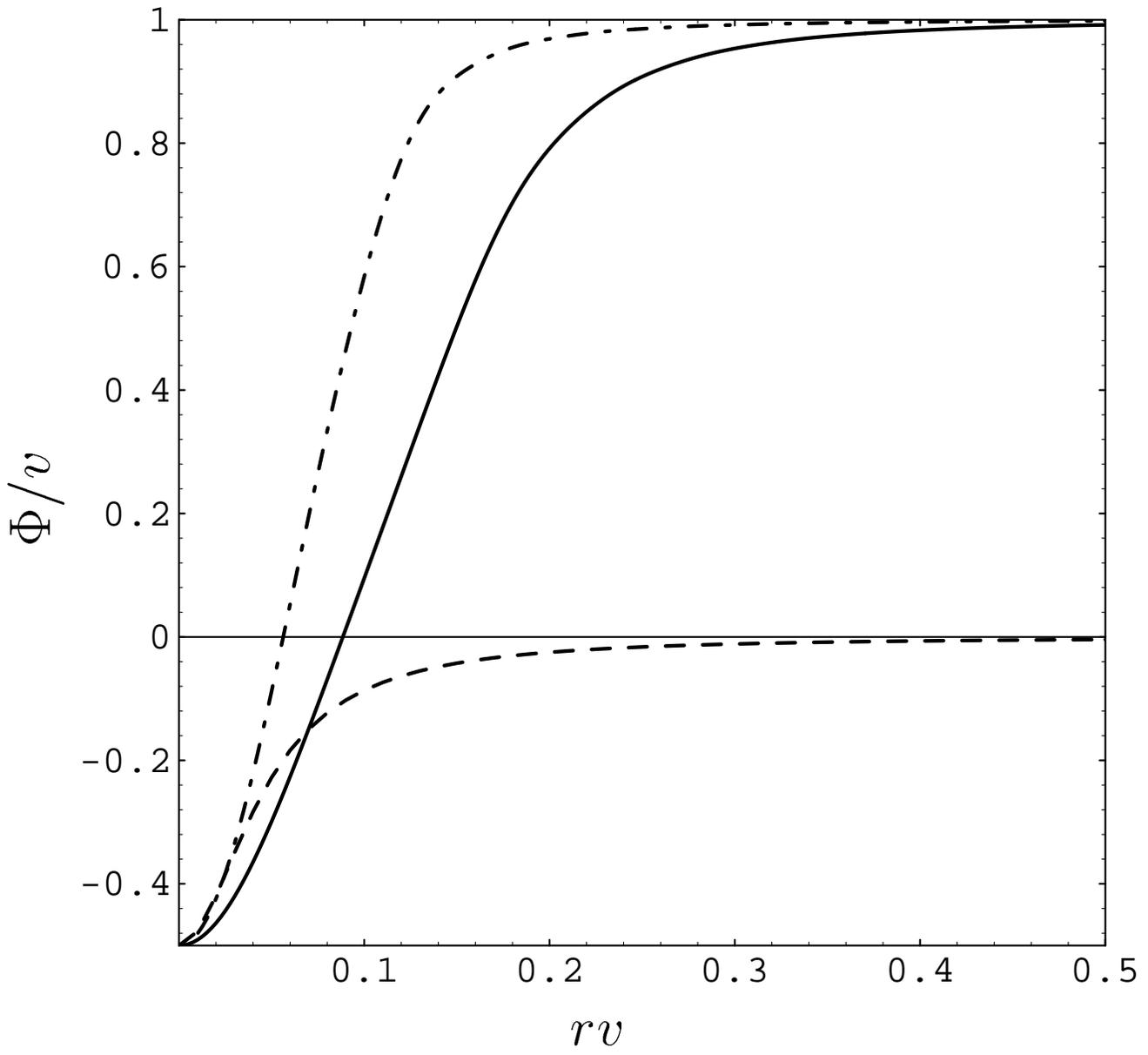
 \item{Fig.2}
      ~~ Numerical Solutions for the Higgs-field in the imaginary Yukawa 
      system. A solid line and dash-dotted line correspond to
      solutions with the Yukawa coupling $|g|=175$ GeV$/v$ for
      $M_h=60$ GeV and 120 GeV 
      respectively.
      A dashed line denotes Inagaki's analytic solution with $v$ = 0.
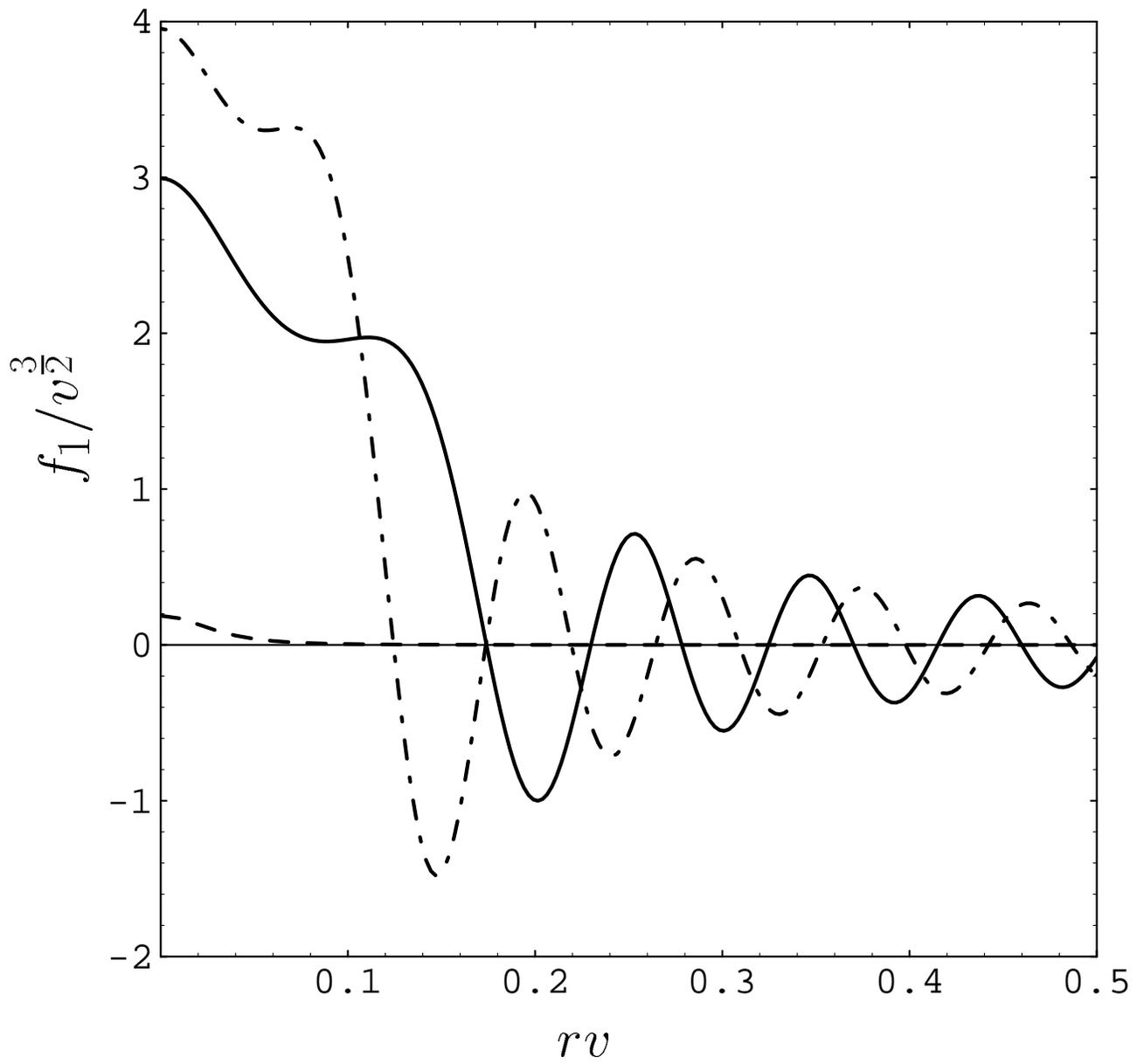
 \item{Fig.3}
      ~~Numerical Solutions for the $f_1$ in the imaginary Yukawa
      system. Each line corresponds to the same one in  Fig.2.
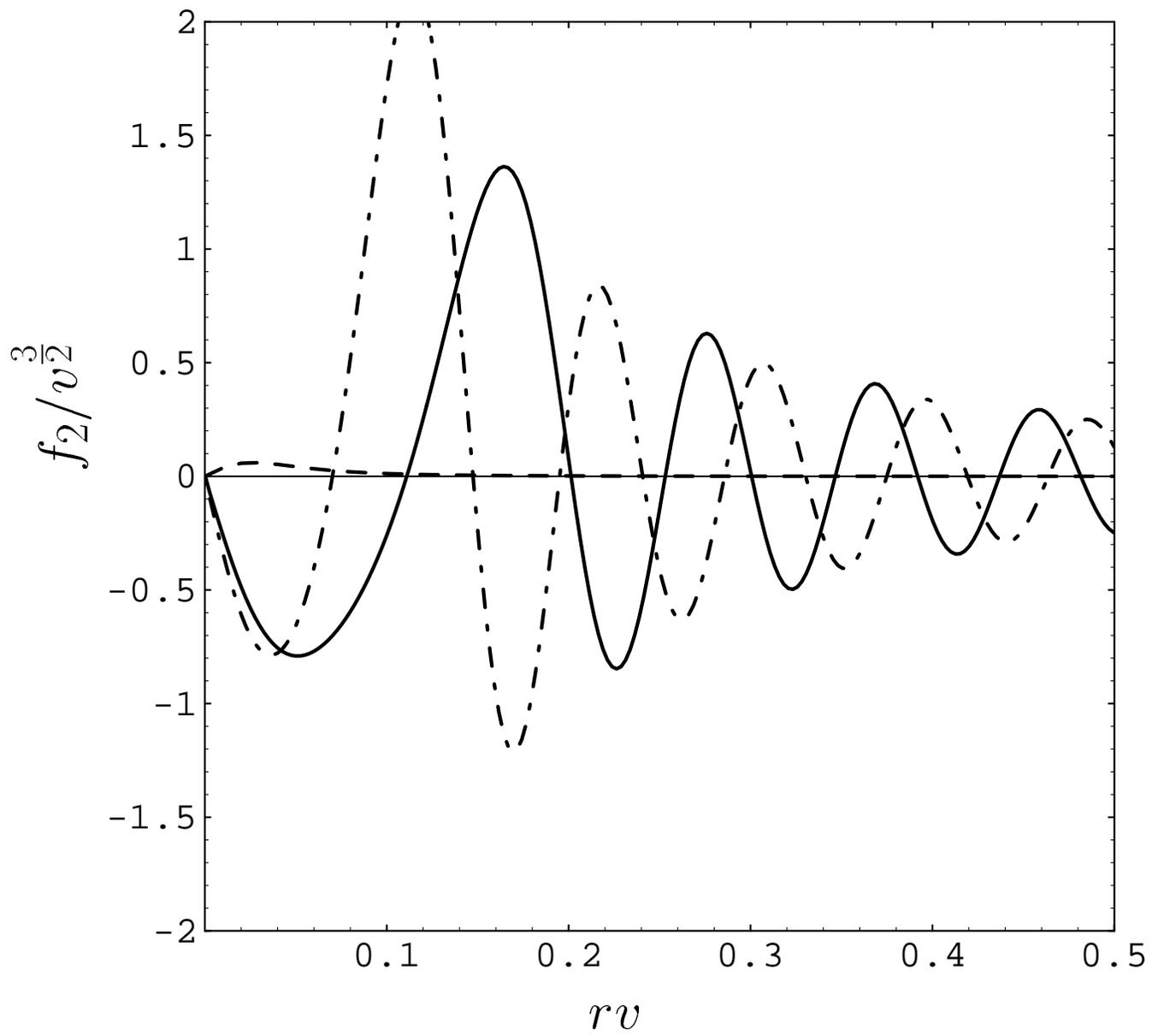
 \item{Fig.4}
      ~~Numerical Solutions for the $f_2$ in the imaginary Yukawa system.
      .Each line corresponds to the same one in  Fig.2.
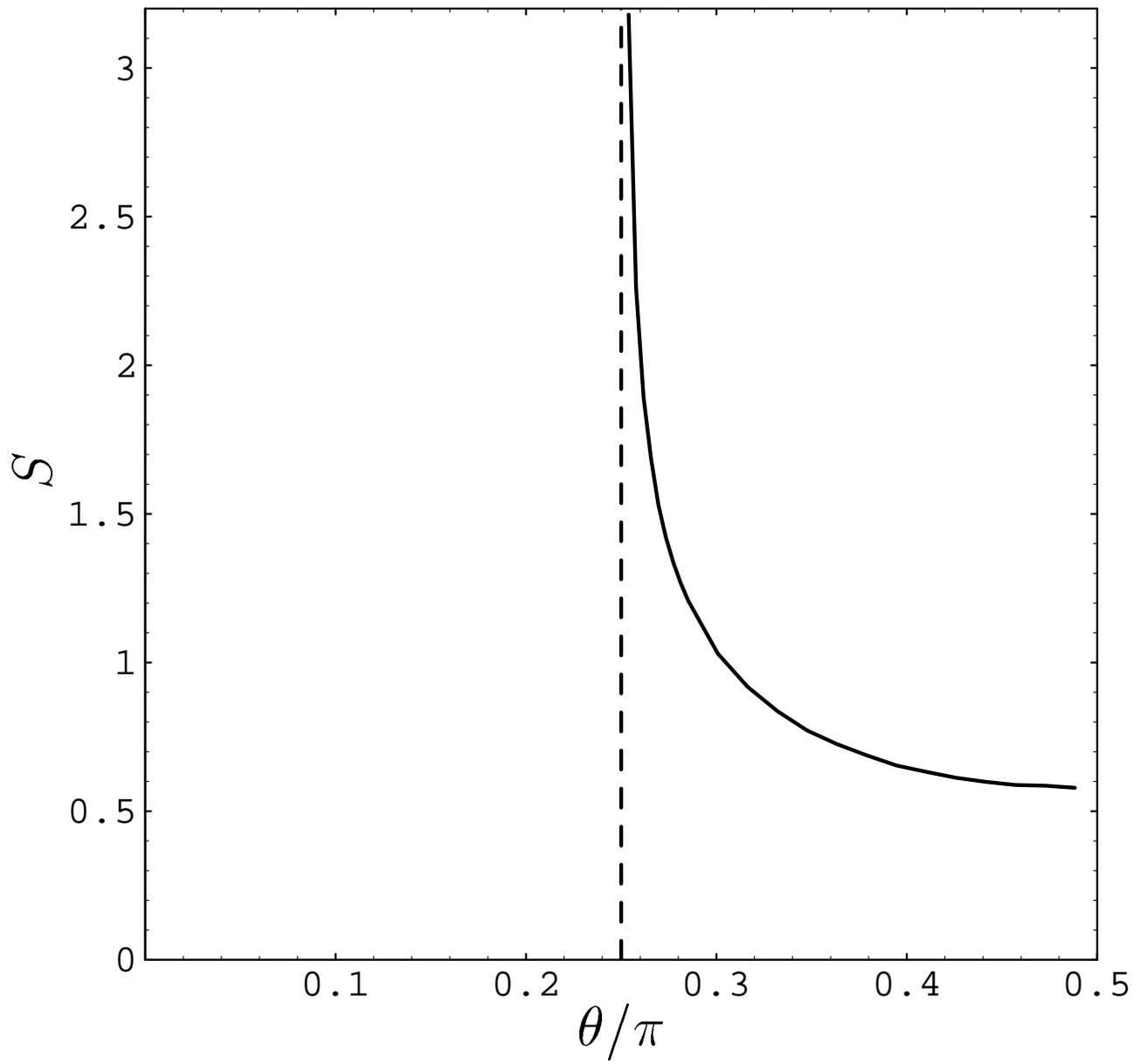
 \item{Fig.5}
      ~~Action ($S$) versus Yukawa coupling phase ($\theta$) in the 
      imaginary Yukawa system.

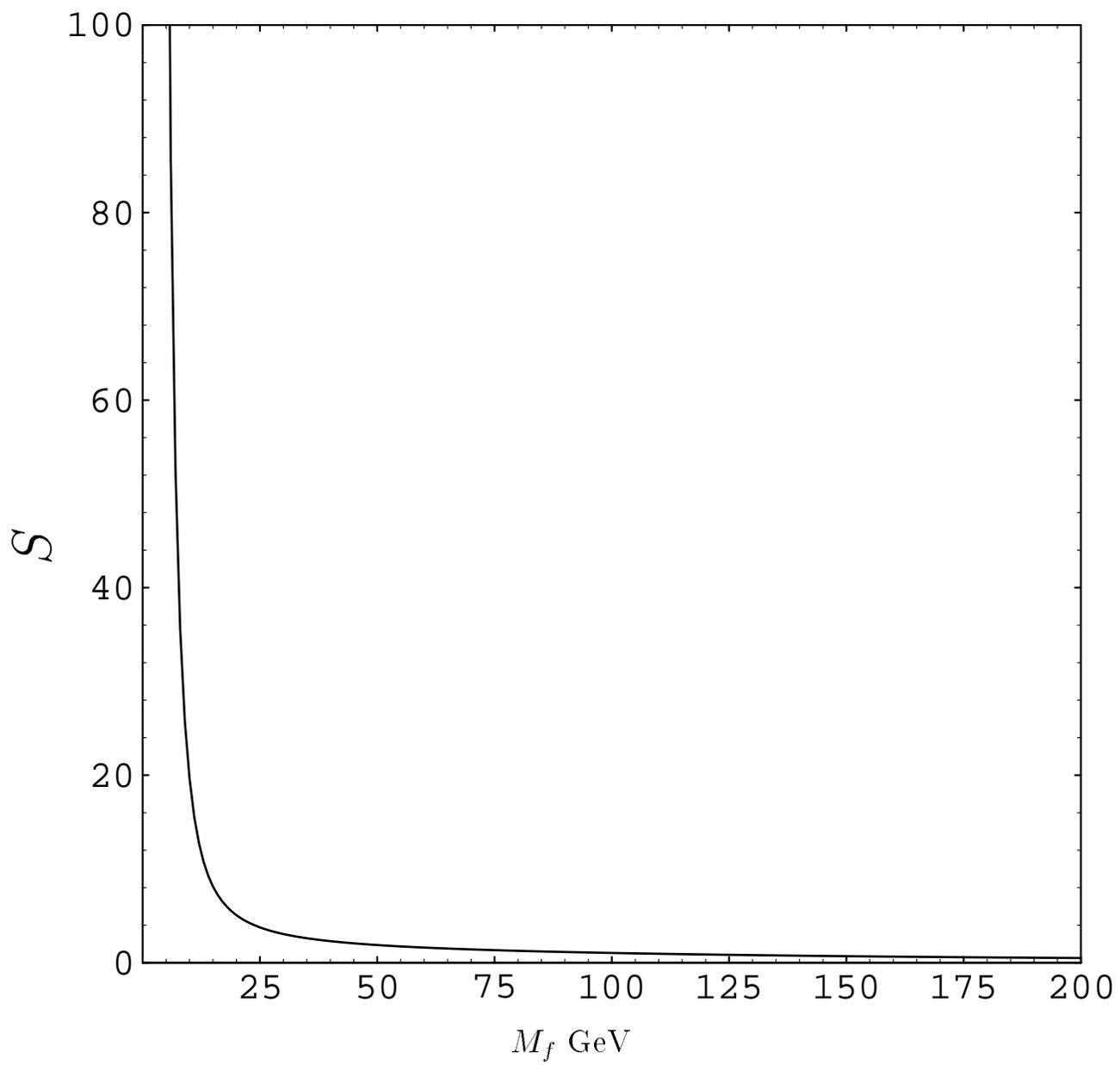
 \item{Fig.6}
      ~~Action ($S$) versus Yukawa coupling strength ($M_f=|g| v$) in the 
      imaginary Yukawa system.

\end{itemize}

\end{document}